\documentclass[a4paper,12pt]{article}

\usepackage{amsmath, amsthm, amsfonts, amssymb}
\usepackage{graphicx}
\usepackage[mathscr]{eucal}

\theoremstyle{plain}
\newtheorem{theorem}{Theorem}[section]

\newtheorem{lemma}{Lemma}[section]

\newtheorem{prop}{Proposition}[section]

\theoremstyle{remark}

\numberwithin{equation}{section}

\def\g{\gamma}
\def\G{\Gamma}
\def\l{\lambda}
\def\p{\partial}
\def\D{\Delta}

\def\k{\varkappa}
\def\Ex{\mbox{\rm e}}
\def\a{\alpha}
\def\b{\beta}
\def\t{\widetilde}
\def\si{\sigma}

\def\d{\delta}

\def\RE{\mathrm{Re}\,}
\def\SS{W_2} 
\def\supp{\mathrm{supp}\,}

\newcommand{\QED}{\mbox{\rule[-1.5pt]{6pt}{10pt}}}

\newcommand{\PF}[1]
{\noindent\textbf{#1}}

\begin{document}

\title{\textbf{Exponential splitting of bound states in
a waveguide with a pair of distant windows}}
\author{D.~Borisov$^a$ and P.~Exner$^{b,c}$}
\date{}
\maketitle

\begin{quote}
{\small {\em a) Bashkir State Pedagogical University, October
Revolution
\\
\phantom{a) } St.~3a, 450000 Ufa, Russia
\\
b) Nuclear Physics Institute, Academy of Sciences, 25068 \v Re\v
z
\\
\phantom{a) }near Prague, Czechia
\\
c) Doppler Institute, Czech Technical University, B\v rehov{\'a}
7,
\\
\phantom{a) }11519 Prague, Czechia
}
\\
\phantom{a) }\texttt{BorisovDI@ic.bashedu.ru},
\texttt{exner@ujf.cas.cz}}
\end{quote}

\begin{quote}
{\small We consider Laplacian in a straight planar strip with
Dirichlet boundary which has two Neumann ``windows'' of the same
length the centers of which are $2l$ apart, and study the
asymptotic behaviour of the discrete spectrum as $l\to\infty$. It
is shown that there are pairs of eigenvalues around each isolated
eigenvalue of a single-window strip and their distances vanish
exponentially in the limit $l\to\infty$. We derive an asymptotic
expansion also in the case where a single window gives rise to a
threshold resonance which the presence of the other window turns
into a single isolated eigenvalue.}
\end{quote}

\setcounter{equation}{0}
\section{Introduction}

Geometrically induced bound states in waveguide systems have
attracted a lot of attention recently. The main reason is that
they represent an interesting physical effect with important
applications in nanophysical devices, but also in flat
electromagnetic waveguides -- cf.~\cite{LCM} and references
therein. At the same time, such a discrete spectrum poses many
interesting mathematical questions.

One of the simplest systems of this kind is a straight hard-wall
strip in the plane with a ``window'' or several ``windows'' in its
boundary modeled by switching the Dirichlet boundary condition to
Neumann in the Laplace operator which will be the Hamiltonian of
our system. By an easy symmetry argument it represents the
nontrivial part of the problem for a pair of adjacent parallel
waveguides coupled by a window or several windows in the common
boundary \cite{ESTV}; this explains the name we use for the
Neumann segments.

The discrete spectrum of such a system is nonempty once a Neumann
window is present. Various properties of these bound states were
analyzed including their number and behaviour with respect to
parameters. Recently we discussed the way in which the eigenvalues
emerge from the continuous spectrum as the window width is
increasing \cite{BEG}; we refer to this paper for references to
earlier work. Here we address a different question: we consider a
strip with a pair of identical Neumann windows at the same side of
the boundary and ask about the behaviour of the discrete spectrum
as the distance between them grows.

There is a natural analogy with the multiple-well problem in the
usual Schr\"odin\-ger operator theory -- see \cite{BCD} and
references therein or \cite[Sec.~8.6]{Da} -- even if the nature of
the effect is different. Recall that in waveguides of the
considered type there are no classically closed trajectories apart
of the trivial set of measure zero, and likewise, there are no
classically forbidden regions. Hence the semiclassical analysis
does not apply here, in particular, there is no Agmon metric to
gauge the distance of the windows which replace potential wells in
our situation.

Nevertheless, the picture we obtain is similar to double-well
Schr\"odin\-ger operators. If the half-distance $l$ between the
windows is large, there is pair of eigenvalues, above and below
each isolated eigenvalue of the corresponding single-window strip.
We will derive an asymptotic expansion which shows that the pair
splitting vanishes exponentially as $l\to\infty$ together with the
appropriate expansion for the eigenfunctions. On the other hand,
the analogy a double-well Schr\"odin\-ger operator can be
misleading. This is illustrated by the case when the single-window
strip has a threshold resonance, which turns into a (single)
isolated eigenvalue under influence of the other window. We derive
the asymptotic expansion as $l\to\infty$ for this case too; it
appears that it is exponential again with the power determined by
the term coming from the second transverse mode present in the
expansion of the resonance wavefunction.

Let us describe briefly the contents of the paper. In the next
section we formulate the problem precisely and state two theorems
which express our main results. In Section~3 we collect general
properties of the involved operator. Before coming to the proper
proofs, we analyze in Sections~4 and 5 strip with a single window,
in particular, we show how the original question stated in PDE
terms can be reformulated as a pair Fredholm problem, the second
being obtained from the first one as a perturbation. Finally, in
Sections~6 and 7 we prove Thms.~\ref{th1} and \ref{th3}.

\setcounter{equation}{0}
\section{Formulation of the problem and the \\ main results}

\begin{figure}
\begin{center}
\noindent\includegraphics[height=5.4 true cm, width=11.94 true
cm]{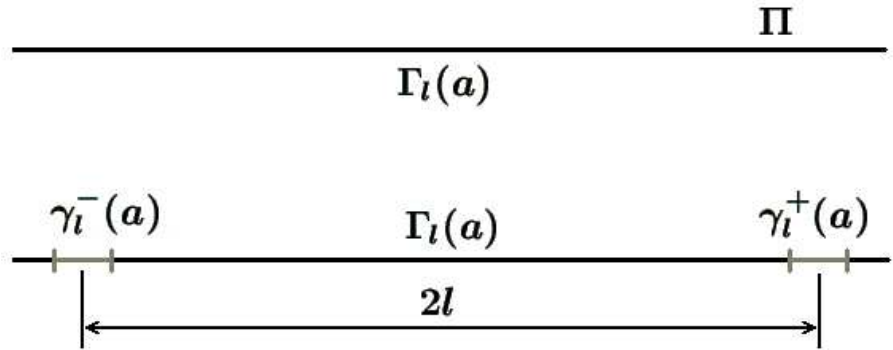}

\medskip

\caption{Waveguide with two Neumann segments}
\label{twowindows}
\end{center}
\end{figure}

Let $x=(x_1,x_2)$ be Cartesian coordinates and suppose that $\Pi$
is a horizontal strip of a width $d$, i.e. $\Pi:=\{x: 0<x_2<d\}$.
In the lower boundary of the strip we select two segments of the
same length $2a$. The distance between these segments, denoted as
$2l$, will be large playing the role of parameter in our
asymptotic expansions. We will employ the symbol $\g_l(a)$ for the
union of these segments, $\g_l(a):=\g_l^+(a)\cup\g_l^-(a)$, where
$\g_l^\pm(a)=\{x: |x_1\mp l|<a,\, x_2=0\}$. The remaining part of
the boundary of $\Pi$ will be indicated by $\G_l(a)$ (cf.
Figure~\ref{twowindows}). The main object of our interest are
discrete eigenvalues of the Laplacian in $\Pi$ with Dirichlet
boundary condition on $\G_l(a)$ and Neumann one on $\g_l$. We
denote such an operator by $H_l(a)$ and look what happens if
$l\to\infty$.

In order to formulate main results of this paper we need some
additional notations and preliminary results concerning a
single-window strip. Denote $\g(a):=\{x: |x_1|<a,\, x_2=0\}$,
where $\G(a):=\p\Pi\setminus\g(a)$. It was proven in \cite{ESTV}
that the Laplacian in $\Pi$ with Dirichlet condition on $\G(a)$
and Neumann one on $\g(a)$ has (simple) eigenvalues below the
threshold of the continuous spectrum for any $a>0$; their number
is finite and depends on $a$. We will indicate the operator in
question and its eigenvalues by $H(a)$ and $\l_j(a)$,
$j=1,\ldots,n$, respectively, with the natural ordering,
$\l_1(a)<\l_2(a)<\ldots<\l_n(a)<\frac{\pi^2}{d^2}$, supposing that
the corresponding eigenfunctions $\psi_j$ are normalized in
$L^2(\Pi)$. Furthermore, it was shown in \cite{ESTV} that there
are critical values of size of Neumann segment,
$0=a_0<a_1<a_2<\cdots$, for which the system has in addition a
threshold resonance, i.e. the equation $(H(a_n)+1)\psi=0$ has a
nontrivial solution $\psi^n(x)$ unique up to a multiplicative
constant. This solution and eigenfunction $\psi_j$ mentioned above
have a definite parity with respect to $x_1$ and behave in the
limit $x_1\to+\infty$ as
\begin{align}\label{1.1}
&\psi^n(x)=\sqrt{\frac{2}{d}}\,\sin\left(\frac{\pi x_2}{d}\right)
+ \b_n\, \Ex^{-\frac{\pi\sqrt{3}}{d}\,x_1}\sin\left(\frac{2\pi
x_2}{d}\right)+
\mathcal{O}\left(\Ex^{-\frac{\pi\sqrt{8}}{d}\,x_1}\right),
\\
&
\psi_j(x)=\a_j\,\Ex^{-\sqrt{\frac{\pi^2}{d^2}-\l_0}\,x_1}\sin\left(\frac{2\pi
x_2}{d}\right)+
\mathcal{O}\left(\Ex^{-\sqrt{\frac{4\pi^2}{d^2}-\l_0}\,x_1}\right),
\label{1.5}
\end{align}
with some constants $\a_j$, $\b_n$; it is clear that
$\a_j=\a_j(a)$. While normalization of $\psi_j$ is natural, the
normalization of $\psi^n$ can be arbitrary, of course; we choose
it in such a way that asymptotically the function coincides with
the first normalized transverse mode. Needless to say, when the
window size is made larger than the critical value, the threshold
resonance turns into a true eigenvalue.

Now we are ready to formulate the main results.
\begin{theorem}\label{th1}
Let the window length be non-critical, i.e. $a\in(a_{n-1},a_n)$
for some $n\in\mathbb{N}$. Then the operator $H_l(a)$ has for
any $l$ large enough exactly $2n$ eigenvalues $\l_j^\pm(l,a)$,
$j=1,\ldots,n$, situated in the interval $(\frac{\pi^2}{4d^2},
\frac{\pi^2}{d^2})$. Each of them is simple and has the
asymptotic expansions
\begin{equation}\label{asm}
\l_j^\pm(l,a)=\l_j(a)\mp\mu_j(a)\,
\mathrm{e}^{-2l\sqrt{\frac{\pi^2}{d^2}-\l_j(a)}}+
\mathcal{O}\left(\mathrm{e}^{-(4\sqrt{\frac{\pi^2}{d^2}
-\l_j(a)}-\si)l}\right)\,,
\end{equation}
as $l\to\infty$ for $j=1,\dots,n$, where $\si$ is an arbitrary
fixed positive number. The coefficient $\mu_j$ is given by
\begin{equation}
\mu_j(a):=\a_j(a)^2d\,\sqrt{\frac{\pi^2}{d^2}-\l_0}\,,\label{ltd1}
\end{equation}
or alternatively by
\begin{equation}
\mu_j(a):=\frac{\pi^2}{d^3\sqrt{\frac{\pi^2}{d^2}-\l_j(a)}}
\left(\int\limits_{\g(a)}\psi_j(x)\,
\mathrm{e}^{\sqrt{\frac{\pi^2}{d^2}-\l_j(a)}\,x_1}\,
dx_1\right)^{\!\!2}.\label{ldt}
\end{equation}
The eigenfunctions $\psi_j^\pm(x)$ associated with eigenvalues
$\l_j^\pm(l,a)$, $j=1,\ldots,n$, have a definite parity being even
for $\l_j^+(l,a)$ and odd for $\l_j^-(l,a)$. Furthermore, in the
halfstrips $\Pi^\pm:=\{x: \pm x_1>0,\: 0<x_2<d\}$ they can be
approximated by
\begin{align*}
&\psi_j^+(x)=\psi_j(x_1\mp l,x_2)+\mathcal{O}\left(
\mathrm{e}^{-(2\sqrt{\frac{\pi^2}{d^2} -\l_j(a)}-\si)l}\right),
\\
&\psi_j^-(x)=\pm\psi_j(x_1\mp l,x_2)+\mathcal{O}\left(
\mathrm{e}^{-(2\sqrt{\frac{\pi^2}{d^2} -\l_j(a)}-\si)l}\right),
\end{align*}
in $\SS^1(\Pi^\pm)$ as $l\to\infty$.
\end{theorem}
\begin{theorem}\label{th3}
Let the Neumann segment have a critical size, $a=a_n$. Then the
operator $H_l(a)$ has $2n+1$ eigenvalues in $(\frac{\pi^2}{4d^2},
\frac{\pi^2}{d^2})$ for $l$ large enough. The first $2n$ of them
together with the associated eigenfunctions behave according to
Theorem~\ref{th1}, while the last one, $\l^+_{n+1}(l,a_n)$,
exhibits the asymptotics
\begin{equation}\label{1.3}
\l^+_{n+1}(l,a_n) =\frac{\pi^2}{d^2}-
\mu\,\Ex^{-\frac{4\sqrt{3}\pi}{d}\,l}
+\mathcal{O}\left(\Ex^{-\frac{2(\sqrt{8}+\sqrt{3})\pi}{d}\,l}\right)\,,
\end{equation}
where
\begin{equation}
\mu:=3\b_n^4 d^2\,,\label{1.4}
\end{equation}
or alternatively,
\begin{equation}\label{1.6}
\mu:=\frac{16}{3d^2}\left(\int\limits_{\g(a_n)}\psi^n(x)\,
\Ex^{\frac{\pi\sqrt{3}}{d}\,x_1}\,dx_1\right)^4.
\end{equation}
The associated eigenfunction $\psi^+_{n+1}$ is even w.r.t. $x_1$
and for any $R$ in the rectangles $\{x: |x_1\mp l|<R\}\cap\Pi$ it
can be approximated for large values of $l$ as
\begin{equation*}
\psi_{n+1}^+(x)= \psi^n(x_1\mp l,x_2)+ \mathcal{O}\left(
\Ex^{-\frac{2\sqrt{3}\pi}{d}\,l}\right)\,
\end{equation*}
in $\SS^1$-norm. In addition it behaves in the limits
$x_1\to\pm\infty$ as
\begin{align*}
\psi^+_{n+1}(x) &=\sqrt{\frac{2}{d}}\, \Ex^{-\varkappa|x_1|}\sin
{\frac{\pi x_2}{d}}+
\mathcal{O}\left(\Ex^{-\frac{\pi\sqrt{3}}{d}|x_1|}\right)\,,
\\ \varkappa & :=\sqrt{\frac{\pi^2}{d^2}-\l_{n+1}}
=\sqrt{\mu}\,\Ex^{-2\frac{\sqrt{3}\pi}{d}l}+
\mathcal{O}\left(\Ex^{-\frac{2\sqrt{8}\pi}{d}l}\right)\,.
\end{align*}
\end{theorem}
\vspace{1em}

\noindent Before proceeding further, let us recall what we have
said in the introduction about the analogy with the multi-well
problem for Schr\"odinger operators. As in that case a (simple)
eigenvalue of the single window problem gives rise to a pair of
eigenvalues (corresponding to eigenfunctions of different
parities) which are exponentially close to each other with respect
to the window distance, and moreover, in the generic case the
splitting is determined by the eigenvalue distance from the
threshold. At a glance the multiplicity is doubled by the
perturbation, however, in reality the problem decomposes due to
mirror symmetry into a pair of problems with definite parities
whose eigenvalues tend to the same limit (see below and Sec. 5).
On the other hand, the asymptotics (\ref{1.3}) in the critical
case differs from what the Schr\"odinger operator analogy would
suggest being determined by the distance from the second
transverse eigenvalue.

Let us now describe our way to prove Theorems~\ref{th1} and
\ref{th3}. The main idea is to reduce the eigenvalue problem at
hand to a Fredholm operator equation of the second kind with a
regular perturbation. Investigating this problem we will get the
result both in the generic situation described in
Theorem~\ref{th1} and for perturbation of a threshold resonance,
just the analysis in the latter case is more subtle.

Our task can be simplified by taking into account the symmetry of
the problem with respect to reflections, $x_1\to -x_1$, which
means that the operator decomposes into orthogonal sum of parts of
a definite parity which can be considered separately. This allows
us to cut the strip $\Pi$ into a pair of halfstrips $\Pi^\pm$ and
to consider the Laplacian in $\Pi^+$ with Dirichlet condition
everywhere at the horizontal boundaries of the halfstrip except
for $\g^+_l(a)$, where the boundary condition is Neumann.
According to the chosen parity of an eigenfunction $\psi$ we
impose at that Dirichlet condition for odd eigenfunctions of the
original problem at the vertical part of the boundary, $x_1=0$, or
Neumann for the even ones. Moreover, it is convenient to shift the
halfstrip by $x_1\to x_1-l$ in order to fix position of the
Neumann segment of the boundary. As a result, we arrive at the
following  pair of eigenvalues problems,
\begin{equation}\label{1.2}
\begin{aligned}
{}&-\D\psi=\l\psi\,,\quad x\in\Pi^l\,, \\ \psi=0\,,\;\;
x\in\G(a)\,,\quad\; {}&\frac{\p\psi}{\p x_2}=0\,,\;\;
x\in\g(a)\,,\quad\; hu=0\,,\;\; x_1=-l\,.
\end{aligned}
\end{equation}
Here $\Pi^l:=\{x\in\Pi:\: x_1>-l\}$ is the shifted halfstrip and
$h$ is the boundary operator which acts as $h u=u$ or $h
u=\frac{\p u}{\p x_1}$ in the odd and even case, respectively.
Eigenvalues of (\ref{1.2}) obviously coincide with those of
$H_l(a)$ and by the even/odd extension one gets the
eigenfunctions of the original problem.

Finally, we remark that
the problem has a simple behaviour with respect to scaling
transformations which allows us to perform the proofs for
$d=\pi$ only.

\section{Preliminaries}

Let us collect first some general properties of the spectrum of
our operators.

\begin{prop}\label{lmpr}
The discrete spectrum of the operator $H_l(a)$ is non-empty for
any $l>a>0$. It consists of a finite number of simple eigenvalues
contained in the interval $\left(\frac{1}{4},1 \right)$ for
$d=\pi$ which depend continuously on $l$ and $a$; for a fixed $a$
those corresponding to even and odd eigenfunctions are increasing
and decreasing, respectively, as functions of the window
separation parameter $l$. All eigenvalues of $H_l(a)$ which remain
separated from the continuum converge to those of $H(a)$ as
$l\to+\infty$, and to each eigenvalue of $H(a)$ there exists a
pair of eigenvalues of $H_l(a)$ associated with eigenfunctions of
opposite parities converging to that eigenvalue of $H_l(a)$. If
the Neumann segment has a critical width, $a=a_n$, then there is a
unique eigenvalue (corresponding to an even eigenfunction) which
tends to one as $l\to+\infty$.
\end{prop}
\PF{Proof.} By the minimax principle and an elementary bracketing
estimate the eigenvalues of $H_l(a)$ can be squeezed between those
of $H(l+a)$ and $H(a)$. The essential spectrum of all the three
operators is the same being equal to $[1,\infty)$; this fact in
combination with the results of \cite{ESTV} shows that
$\sigma_\mathrm{disc} (H_l(a))$ is non-empty, finite, and
contained in $\left(\frac{1}{4},1 \right)$. A similar bracketing
argument shows that the eigenvalues $\l_j^\pm(l,a)$ of the problem
(\ref{1.2}) with Neumann and Dirichlet boundary condition at
$x_1=-l$, respectively, satisfy $\l_j^+(l,a) \le \l_j(a) \le
\l_j^-(l,a)$ for $j=1,\dots,n$, where the upper bound is replaced
by one if the Dirichlet problem has less than $j$ eigenvalues. In
fact, bracketing implies also the stated monotonous behaviour with
respect to $l$, i.e.
\begin{equation} \label{DNest}
\l_j^+(l',a) \le \l_j^+(l,a) \le \l_j(a) \le \l_j^-(l,a) \le
\l_j^-(l',a)
\end{equation}
for $l'\ge l$ with the same convention as above; it is sufficient
to write $\Pi_{l'}$ as a union of $\Pi_l$ and a rectangle
separated by an additional Neumann or Dirichlet boundary condition
and to realize that in neither of these cases the rectangle can
contribute to the spectrum below the continuum threshold, because
it has Dirichlet condition at the horizontal part of the boundary.
In addition, the standard domain-changing argument
\cite[Sec.~VII.6.5]{Kt} shows that the functions
$\l_j^\pm(\cdot,a)$ are continuous. In view of the monotonicity
mentioned above their limits as $l\to\infty$ exist; it remains to
check that $\l_j^\pm(l,a)\to \l_j(a)$.

Take $\psi\in D(H(a))$ and a function $g\in C_0^\infty$ such that
$g(x)=0$ for $x\le 0$ and $g(x)=1$ for $x\ge 1$. Denoting
$h_l(x,y):= g(2(x+l)/l)$ we can construct a family $\{\psi_l\}$ by
$\psi_l(x,y):= \psi(x,y)h_l(x,y)$; by construction the function
$\psi_l$ belongs to the domain of $H^-_l(a)$ which is the
Laplacian with the boundary condition as in (\ref{1.2}) for
$hu=u$. Using the fact that $\|\nabla h_l\|^2= 2\|g'\|^2 l^{-1}$
and $\|\Delta h_l\|^2= 8\|g''\|^2 l^{-3}$ one can check easily
that $\psi_l\to \psi$ and $H^-_l(a)\psi_l \to H(a)\psi$ as $l\to
\infty$, so $H^-_l(a)\to H(a)$ in the strong-graph sense. By
\cite[Thm~VIII.26]{RS} this is equivalent to the strong resolvent
convergence, hence to each $\l_j(a)$ there is a family of
$\l^-_j(a)$ converging to that value. Since the spectrum of
$H_l(a)$ in $(\frac{1}{4},1)$ is discrete, simple, finite, and
depends monotonously on $l$, we get the desired result. In a
similar way one can check that $\l_j^+(l,a)\to \l_j(a)$ as
$l\to\infty$.

The continuity w.r.t. $a$ is proved as the $a$-continuity in case
of a single Neumann window. We expand the solution inside and
outside the window regions with respect to the appropriate
transverse bases and match the Ans\"atze smoothly at the window
edges. This yields an infinite family of linear equations for the
coefficients of the expansions, which can be regarded as a search
for the kernel of a certain operator in the $\ell^2$ space of the
coefficients with a properly chosen weight. One has to check that
this operator is Hilbert-Schmidt and continuous with respect to
the parameters in the Hilbert-Schmidt norm. The argument is
analogous to that from the proof of Proposition~2.1 of \cite{BEG},
so we skip the details; the only difference is that due to the
lack of symmetry the matching has to be performed at each window
separately and the coefficient space is ``twice as large''.

It remains to check the last claim. Using bracketing once more we
see that if the presence of the other window turns a threshold
resonance into an eigenvalue, the corresponding eigenfunction must
be symmetric; in view of the proved monotonicity it is sufficient
to show that this happens for $l$ large enough. Since this part of
the proposition is not used in the proof of the claim of
Theorem~\ref{th3} concerning the first $2n$ eigenvalues
$\l_j^\pm$, we may assume that it is already proven and that we
thus know that for large $l$ the operator $H_l(a_n)$ posesses $2n$
eigenvalues $\l_j^\pm \in \left(\frac{1}{4},1 \right)$
corresponding to eigenfunctions $\psi_j^\pm$. We seek a
$(2n+1)$-dimensional subspace such that for any $\psi$ from it we
have $(\psi,H_l(a_n)\psi)- \|\psi\|^2<0$. To this aim we employ a
Goldstone-Jaffe-type argument inspired by \cite{ESTV} and choose
\begin{equation*}
\psi= c_0(\chi_{L,\varsigma}\psi^n + \varepsilon p)+ \sum_{j=1}^n
c_j^\pm \psi_j^\pm\,,
\end{equation*}
where $\psi^n$ is the resonance function (\ref{1.1}),
$\chi_{L,\varsigma}:\mathbb{R}\to (0,1]$ equals one in $(-L,L)$
for some $L>l+a$ and $\chi_{L,\varsigma}= \exp(-\varsigma(|x|-L))$
otherwise, and $p$ is a $C_0^\infty$ function supported in the
other window region. In view of the asymptotic behaviour
(\ref{1.1}) such functions span a subspace of the needed
dimension. Evaluating the energy form $(\psi,H_l(a_n)\psi)-
\|\psi\|^2$ we see that if some of the coefficients $c_j^\pm$ is
nonzero, it is negative even with $\varepsilon=0$. In the opposite
case we use the fact that in the leading term we have, as in
\cite{ESTV}, two competing terms, one linear in $\varepsilon$ and
the other positive coming from the tails of $\psi$ controlled by
the parameters $L$ and $\varsigma$; we can choose them in such a
way that the form is negative again.

Hence $H_l(a_n)$ has for $l$ sufficiently large at least $2n+1$
eigenvalues. In fact, it has exactly this number, because its
symmetric and antisymmetric parts have for large $l$ enough $l+1$
and $l$ eigenvalues, respectively, otherwise we would have an
contradiction with the monotonicity and continuity properties
stated above. In particular, the largest eigenvalue is increasing
w.r.t. $l$ since it corresponds to an even eigenfunction. In view
of (\ref{DNest}) and the fact that $a_n$ is the critical width, we
conclude that $\l_{n+1}^+\to 1-$ as $l\to\infty$. \quad \QED

\section{Analysis of the limiting operator}

After these preliminaries let us pass to the proper subject of the
paper. First we are going to discuss the limiting, i.e. one-window
operator which means to analyze the following boundary value
problem,
\begin{equation}\label{2.1}
-(\D+\l) u=f\,,\;\; x\in\Pi\,,\qquad u=0\,,\;\, x\in\G(a)\,,
\quad\; \frac{\p u}{\p x_2}=0\,,\;\, x\in\g(a)\,.
\end{equation}
The right hand side $f$ is here assumed to be finite and to belong
to $L^2(\Pi)$; our aim is to discuss the existence and uniqueness
of the solution to (\ref{2.1}) as well as its dependence on $\l$.
The method we use is to reduce (\ref{2.1}) to a Fredholm operator
equation. Then our task will be reduced to analysis of operator
families, in particular their holomorphic dependence on the
spectral parameter $\l$ (one need not specify at that the topology
-- cf.~\cite[Sec.~VI.3]{RS}). The reduction will follow a general
scheme proposed by Sanchez-Palencia \cite{SP} and it will be
analogous to the treatment of a similar problem in
\cite[Sec.~3.1]{BEG}.

We will use the symbol $\mathcal{D}_\d$ to indicate the open
subset $\{\l: \RE\l<\d\}$ of the complex plane. The structure of
the solution to the problem (\ref{2.1}) for $\l$ close to $1$ and
for $\l$ separated from $1$ is different. This is the reason why
we will consider these two cases separately. We suppose first that
$\l\in\mathcal{D}_\d$, where $\l_n(a)<\d<1$. In this situation it
is sufficient to consider solutions of the problem (\ref{2.1}) in
the class of functions which behave as $\mathcal{O}
(\Ex^{-\sqrt{1-\l}|x_1|})$ in the limit $|x_1|\to\infty$.

Since the function $f$ is finite by assumption, its support lies
inside the rectangle $\Pi_b:=\Pi\cap\{x: |x_1|<b\}$ for some
$b>0$. Consider two boundary value problems,
\begin{equation}
-(\D+\l) v^\pm=g\,,\quad x\in\Pi^\pm\,,\qquad v^\pm=0\,,\quad
x\in\p\Pi^\pm\,, \label{2.2}
\end{equation}
where $g$ is an arbitrary function from $L^2(\Pi)$ with the
support contained in $\Pi_A$ for some $A\ge\max\{a,b-1\}$. This
choice is given by the requirement that $\Pi_A$ contains both the
window and the support of $f$ in such a way which will make the
smooth interpolation (\ref{2.5}) used below possible. The problems
(\ref{2.2}) can be easily solved by separation of the variables;
using the explicit form of Green's function of Laplace-Dirichlet
problem on a halfline we get
\begin{gather}
v^\pm(x)=\int\limits_{\Pi^\pm} G^\pm(x,t,\l)\,g(t)\,
d^2t\,,\label{2.3} \\
G^\pm(x,t,\l)=\sum\limits_{j=1}^\infty\frac{1}{\pi
\kappa_j(\l)}\left(\Ex^{-\kappa_j(\l)|x_1-t_1|}-\Ex^{\mp
\kappa_j(\l)(x_1+t_1)}\right)\, \sin jx_2\, \sin jt_2\,,
\label{2.12}
\end{gather}
where $\kappa_j(\l):=\sqrt{j^2-\l}\,$. In the following we will
also employ the ``glued'' function $v$ equal to $v^+$ if $x_1\ge
0$ and to $v^-$ if $x_1<0$. The functions $v^\pm$ can be naturally
regarded as results of action of the bounded linear operators
$T_1^\pm(\l)$, i.e. we have $v^\pm=T_1^\pm(\l)g$, where $T^\pm_1:
L^2(\Pi^\pm_A)\to \SS^2(\Pi^\pm)$ with the ``halved'' rectangles
$\Pi^\pm_A=\Pi\cap\{x: 0<\pm x_1<A\}$. It is easy to check that
the operator families $T^\pm_1$ are holomorphic in $\l\in
\mathcal{D}_\d$. In the next step we consider the problem
\begin{equation}\label{2.4}
\D w=\D v\,,\;\; x\in\Pi_A\,,\quad\; \frac{\p w}{\p x_2}=0\,,\;\;
x\in\g(a)\,,\quad  w=v\,,\;\; x\in\p\Pi_A\setminus\g(a)\,.
\end{equation}
The function $v$ may have according to its definition given above
a weak discontinuity, i.e. a jump of the first derivatives. Thus
we have to say what we mean by $\D v$ in (\ref{2.4}): it is the
function from  $L^2(\Pi)$ which coincides with $\D v^+$ if $x_1>0$
and with $\D v^-$ if $x_1<0$. With the problem (\ref{2.2}) in mind
we can also write $\D v=-(\l v+g)$. The problem (\ref{2.4}) is
posed in a bounded domain, hence the standard theory of elliptic
boundary value problems is applicable. In particular, we can infer
using \cite{Ld} that the function $w$ exists, it is unique and
belongs to $\SS^1(\Pi_A)$. We will also consider its restriction
avoiding the points where the boundary condition changes, regarded
as an element of $\SS^1(\Pi_A)\cap\SS^2(\Pi_{A}\setminus S_r)$ for
any $r>0$, where $S_r=\{x: (x_1\pm a)^2+x^2_2<r^2\}$. In this way
we introduce a linear bounded operator $T_2: L^2(\Pi_A)\to
\SS^1(\Pi_A)\cap\SS^2(\Pi_{A}\setminus S_r)$ (for any $r$) such
that $w=T_2g$.

Next we employ a smooth interpolation. Let $\chi$ be an
infinitely differentiable mollifier function such that
$\chi(\tau)=1$ if $|\tau|<A-1$ while for $|\tau|>A$ it vanishes.
We will construct a solution to the problem (\ref{2.1})
interpolating between the functions $v$ and $w$, specifically
\begin{equation}\label{2.5}
u(x)=\chi(x_1)w(x)+(1-\chi(x_1))v(x).
\end{equation}
Since $w=T_2 g$ and $v^\pm=T_1^\pm(\l) g$, we can also regard
$u$ as the result of an action of some linear operator $T_3(\l)$
which maps $L^2(\Pi_A)$ into $\SS^1(\Pi)\cap\SS^2(\Pi\setminus
S_r)$ for a fixed $r>0$. Such an operator $T_3$ is linear and
bounded, and as an operator family with respect to $\l$ it is
again holomorphic.

Owing to the definition of $w$ and $v$ the function $u$
satisfies all the boundary conditions involved in (\ref{2.1}),
and consequently, it represents a solution to (\ref{2.1}) if and
only if it satisfies the differential equation in question.
Substituting (\ref{2.5}) into the latter and taking into account
(\ref{2.2}), (\ref{2.4}), we arrive at the equation
\begin{equation}\label{2.6}
g+T_4(\l)g=f\,,
\end{equation}
where $T_4: L^2(\Pi_A)\to L^2(\Pi_A)$ is a linear bounded
operator defined by
\begin{equation}\label{2.7}
T_4(\l)g:=-2\nabla_x \chi \cdot\nabla_x (w-v) -(w-v)(\D+\l)\chi\,,
\end{equation}
where the dot in the first term denotes the inner product in
$\mathbb{R}^2$. The relation (\ref{2.6}) is the sought Fredholm
equation, considered in the space $L^2(\Pi_A)$. Naturally the
first thing to do here is to check the compactness of the operator
$T_4$. It can be done as follows. The function $w-v$ belongs to
$\SS^1(\Pi_A)$, thus the operator mapping $g$ into $w-v$ is
bounded as an operator from $L^2(\Pi_A)$ into $\SS^1(\Pi_A)$, and
consequently, it is compact as an operator in the space
$L^2(\Pi_A)$; this solves the question for the second term at the
right-hand side of (\ref{2.7}). Furthermore, due to the definition
of the mollifier $\chi$ the support of $\nabla_x \chi$ lies within
$\overline{\Pi}_A\setminus\Pi_{A-1}$. This domain does not contain
the endpoints of the segment $\g$. Hence $w-v\in\SS^2(\supp{
\nabla_x \chi})$, and therefore $\nabla_x (w-v)$ considered as an
element of $L^2(\supp{\nabla_x \chi})$ results from action of a
compact operator mapping $L^2(\Pi_A)$ into $L^2(\supp{\nabla_x
\chi})$; this concludes the proof of compactness $T_4(\l)$
considered as an operator in the space $L^2(\Pi_A)$. In a similar
way one can check that $T_4(\l)$ is a holomorphic operator family
w.r.t. $\l$.

This conclusion allows us to apply to (\ref{2.6}) the standard
Fredholm technique; we will see that solution to (\ref{2.6})
exists and is unique for almost all $\l$ except for points where a
nontrivial solution for (\ref{2.6}) with zero right-hand side
exists. This will yield a solution to our original problem because
the two are equivalent; this is the contents of the following
lemma the proof of which we skip because it is completely
analogous to that of Proposition~3.2  in \cite{BEG}.

\begin{lemma}\label{lm2.1}
To any solution $g$ of (\ref{2.6}) there is a unique solution
$u=T_3(\l)g$ of (\ref{2.1}), and vice versa, for each solution of
(\ref{2.1}) there exists a unique $g$ solving (\ref{2.6}) such
that $u=T_3(\l)g$. The equivalence holds for any $\l\in
\mathcal{D}_\d$.
\end{lemma}

\noindent Thus the equation (\ref{2.6}) says how to find a bounded
solution to (\ref{2.1}): one should solve the equation (\ref{2.7})
and then to construct the solution of (\ref{2.1}) by the procedure
described above, i.e. by putting $u=T_3(\l)g$.

Since the operator family $T_4(\l)$ is holomorphic, the
corresponding resolvent family $(I+T_4(\l))^{-1}$ is meromorphic
and its only poles are exactly the eigenvalues of $H(a)$ --
cf.~\cite[Chap. 16, Th. 7.1]{SP}. In order to prove
Theorem~\ref{th1}, we need to know more about the behavior of
$(I+T_4(\l))^{-1}$ in the vicinity of these poles.

\begin{lemma}\label{lm2.2} Let $\l_0<1$ be an eigenvalue of
$H(a)$. Then for any $\l$ close enough to $\l_0$ the following
representation is valid,
\begin{equation}\label{2.8}
(I+T_4(\l))^{-1}=\frac{\phi}{\l-\l_0}T_5+T_6(\l)\,,
\end{equation}
where $T_5 f:=-(f,\psi)_{L^2(\Pi)}$ and $T_6: L^2(\Pi_A)\to
L^2(\Pi_A)$ is a bounded linear operator which is holomorphic in
$\l$. Furthermore, $\phi$ is such that $\psi=T_3(\l_0)\phi$, where
$\psi$ is an eigenfunction of $H(a)$ associated with $\l_0$ and
normalized in $L^2(\Pi)$.
\end{lemma}
\PF{Proof.} We assume throughout that $\l\in\mathcal{D}_\d$ lies
in a small neighborhood of $\l_0$ containing no other
eigenvalues of $H(a)$. As we have already mentioned, the
operator family $(I+T_4(\l))^{-1}$ has a pole at $\l_0$. It
means that the vector-valued function $g:\lambda\mapsto
(I+T_4(\l))^{-1}f$ satisfies
\begin{equation}\label{2.9}
g(\lambda) =\frac{g_{-q}}{(\l-\l_0)^q}+\frac{\t
g(\l)}{(\l-\l_0)^{q-1}}\,,
\end{equation}
where $q$ is a positive integer and $\t g$ is holomorphic in $\l$.
Substituting this representation into (\ref{2.6}) and calculating
the coefficients of $(\l-\l_0)^{-q}$ we see that $g_{-q}$ must
satisfy the equation $g_{-q}+T(\l_0)g_{-q}=0$, in other words
$g_{-q}=\phi T_5 f$, where $T_5 f$ is a number depending on $f$.
Together with (\ref{2.9}) this means that the solution to
(\ref{2.1}) associated with $g$, i.e. $u=T_3(\l)g$, can be written
as
\begin{equation}\label{2.10}
u(x,\l)=\frac{T_5 f}{(\l-\l_0)^q}\psi(x)+\frac{\t
u(x,\l)}{(\l-\l_0)^{q-1}}\,,
\end{equation}
where is $\t u$ is holomorphic in $\l$. Due to the definition of
$T_3(\l)$ this formula is valid in the sense of $\SS^1(\Pi)$-norm
as well as in $\SS^2(\Pi\setminus\Pi_A)$. Taking the inner product
of (\ref{2.1}) with $\psi$, using the fact that the latter is an
eigenfunction of $H(a)$, and performing an integration by parts in
$\Pi_R$ with $R$ large enough we find
\begin{equation}\label{2.11}
-\int\limits_{\p\Pi_R}\left(\psi\frac{\p u}{\p\nu}-u\frac{\p
\psi}{\p\nu}\right)+\l_0(u,\psi)_{L^2(\Pi_R)}=(f,\psi)_{L^2(\Pi_R)}+
\l(u,\psi)_{L^2(\Pi_R)}\,.
\end{equation}
The functions $u$ and $\psi$ behave at infinity as
$\mathcal{O}(\Ex^{-|x_1|\sqrt{1-\l}})$ and
$\mathcal{O}(\Ex^{-|x_1|\sqrt{1-\l_0}})$, respectively. With this
fact in mind we can pass to the limit $R\to\infty$ in (\ref{2.11})
for each fixed value of $\l$; this implies the identity
\begin{equation*}
\l_0(u,\psi)_{L^2(\Pi)}=(f,\psi)_{L^2(\Pi)}+
\l(u,\psi)_{L^2(\Pi)}\,.
\end{equation*}
Substituting to it from (\ref{2.10}) and computing the
coefficients at the same powers of $\l\!-\!\l_0$, we see first
that $q=1$, and furthermore, that $T_5 f=-(f,\psi)_{L^2(\Pi)}$.
This completes the proof. \quad \QED
\bigskip

We will also need to know the behavior of the inverse
$(I+T_4(\l))^{-1}$ as $\l\to 1$. For the right hand side $f$ in
(\ref{2.1}) with a definite parity w.r.t. $x_1=0$ it was done in
\cite{BEG}, here we have just to show how to extend this result to
our case. We will assume that $\l$ lies in a small neighborhood of
one and that this neighborhood contains no eigenvalues of $H(a)$.
First of all, however, we should characterize the class of
functions in which we will seek the solution of (\ref{2.1}) in
this case. Instead of $\l$ we introduce another parameter by
setting $\l=1-\kappa^2$, where $\kappa$ lies in a small
neighborhood of zero. The only restriction to the size of this
neighborhood correction is that the associated values of $\l$
should not coincide with eigenvalues of the operator $H(a)$. If
$\kappa$ is real and a solution to the problem (\ref{2.1}) exists,
it is unique and holomorphic in $\kappa$. This fact follows from
the arguments given above, because for such a $\l=1-\kappa^2$ the
equation (\ref{2.6}) is uniquely solvable. The said solution can
be extended to all values of $\kappa$ in the vicinity of zero so
that this extension will be an analytic function of $\kappa$. The
existence of such an extension is guaranteed by the definition of
the functions $v^\pm$ in (\ref{2.3}) where $\kappa_1(\l)$ is
nothing else than $\kappa$ introduced above. We see that the
formulae (\ref{2.3}) are valid not only for real $\kappa$ but also
in a complex neighborhood including $\kappa=0$, because the
kernels (\ref{2.12}) have finite limits as $\kappa=0$, namely
\begin{equation}\label{2.13}
\begin{aligned}
G^\pm(x,t,1)={}&-\frac{1}{\pi}(|x_1-t_1|\mp(x_1+t_1))\sin x_2 \sin
t_2 \\
{}& +\sum\limits_{j=2}^\infty\frac{1}{\pi
\kappa_j(1)}\left(\Ex^{-\kappa_j(1)|x_1-t_1|}-\Ex^{\mp
\kappa_j(1)(x_1+t_1)}\right)\sin jx_2 \sin jt_2\,.
\end{aligned}
\end{equation}
This is why we are able to extend the solution of the problem
(\ref{2.1}) analytically to all values of $\kappa$ in the vicinity
of zero. We should also stress that the function $u$ given by
(\ref{2.5}) decays exponentially at infinity if $\RE \kappa>0$, it
is bounded for $\RE \kappa=0$ and increases exponentially provided
$\RE \kappa<0$.

In this approach all the operators introduced above preserve their
properties when we vary the range of the variables passing from
unbounded domains the cut-off ones treating, for instance,
$T_1^\pm$ as operators mapping $L^2(\Pi_A^\pm)$ into
$\SS^2(\Pi_R^\pm)$ for any $R$. As another notational
simplification we will not introduce an extra symbol for the
composed mapping $\kappa\mapsto T_i(1\!-\!\kappa^2)$ and write
instead just $T_i(\kappa)$.

Mimicking the argument used in the proof of \cite[Thm~3.4]{BEG},
one can check the following claim:

\begin{lemma}\label{lm2.3}
If the Neumann segment of the boundary does not have a critical
size, the operator $(I+T_4(\kappa))^{-1}$ exists and is uniformly
bounded in $\kappa$ in the vicinity of zero. In the opposite case,
i.e. $a=a_n$, we have in a punctured neighborhood of zero the
following representation,
\begin{equation}\label{2.14}
(I+T_4(\kappa))^{-1}=\frac{\phi^n}{\kappa}T_7+T_8(\kappa)\,,
\end{equation}
where $T_7 f:=\frac{1}{2}(f,\psi^n)_{L^2(\Pi)}$ and $T_8:
L^2(\Pi_A)\to L^2(\Pi_A)$ is a bounded linear operator which is
holomorphic in $\kappa$. Furthermore, $\phi^n$ is such that
$\psi^n=T_3(\kappa=0)\phi^n$, where $\psi^n$ solves the equation
$(H(a_n)+1)\psi^n=0$ and behaves at infinity in accordance with
(\ref{1.1}).
\end{lemma}

\section{Analysis of perturbed operator}

The main purpose of this section is to reduce the problem
\begin{equation}\label{3.1}
\begin{aligned}
{}&-(\D+\l)u=f\,,\quad x\in\Pi^l\,, \\
\psi=0\,,\;\; x\in\G(a)\,,\quad\; {}&\frac{\p\psi}{\p x_1}=0\,,
\;\; x\in\g(a)\,,\quad\; hu=0\,,\;\; x_1=0\,, \phantom{AAAA}
\end{aligned}
\end{equation}
to an operator equation similar to (\ref{2.6}). We will show that
the problem (\ref{3.1}) can be reduced to solution of a Fredholm
equation which is a regular perturbation of the equation
(\ref{2.6}). We will start from the case of Dirichlet condition at
the cut $x_1=0$, i.e., $hu=u$. We are going to employ the same
scheme as in previous section and use the same notations unless
stated otherwise.

First we will treat the case $\l\in \mathcal{D}_\d$. In analogy
with (\ref{2.2}) we consider two problems,
\begin{align}
{}&-(\D+\l) v^+_l=g\,,\;\; x\in\Pi^+\,,\quad\; v^+_l=0\,,\;\;
x\in\p\Pi^+\,,\label{3.2}
\\
{}&-(\D+\l) v^-_l=g\,,\;\; x\in\Pi^-_l\,,\quad\; v^-_l=0\,,\;\;
x\in\p\Pi^-_l\,.\label{3.3}
\end{align}
The first one coincides with (\ref{2.2}) for $v^+_l$, while in
(\ref{3.3}) we take into account the perturbation. Consequently,
we have $v^+_l:= v^+$,  where $v^+$ is the function from
(\ref{2.3}). The problem (\ref{3.3}) differs from (\ref{2.2}) but
it can be solved again by separation of variables. It is
convenient to write its solution $v_l^-$ in the following form,
\begin{gather}
v^-_l(x)=v^-(x)+\int\limits_{\Pi^-}
G^-_l(x,t,\l)\,g(t)\,d^2t\,,\label{3.4}
\\
G^-_l(x,t,\l)=-\sum\limits_{j=1}^\infty\frac{2\,\Ex^{-\kappa_j(\l)
l}}{\pi \kappa_j(\l)\sinh \kappa_j(\l) l}\,\sinh \kappa_j(\l)
x_1\,\sinh \kappa_j(\l) t_1 \sin j x_2\,\sin j t_2\,,\label{3.7}
\end{gather}
where $v^-$ is given by (\ref{2.3}); we keep in mind here that $g$
is finite, and therefore its support lies inside $\Pi_l$ for all
$l$ large enough. As in previous section we can introduce a linear
bounded operator $T_{9}(\l): L^2(\Pi_A^-)\to \SS^2(\Pi^-_l)$ such
that $v^-_l=T_{9}(\l,l) g$. This operator can be represented as
the sum $T_{9}(\l,l)=T_1^-(\l)+T_{10}(\l,l)$, where $T_{10}(\l,l):
L^2(\Pi_A^-)\to \SS^2(\Pi^-_l)$ is holomorphic in $\l$, jointly
continuous with respect to $(\l,l)$ provided
$\l\in\mathcal{D}_\d$, $l\in[l_0,+\infty]$, and $l_0$ is a fixed
number large enough. The norm of the operator $T_{10}$ is of order
$\mathcal{O}(\Ex^{-l\sqrt{1-\l}})$ as $l\to+\infty$, hence for
$\l\in\mathcal{D}_\d$ we may consider this operator as an
exponentially small perturbation.

The analogue of the function $w$ (denoted here by $w_l$) is
defined as above without any changes, i.e. as a solution of the
problem (\ref{2.4}) with $v$ replaced by
\begin{equation*}
v_l:=\left\{
\begin{aligned}
{}&v^+_l\,,\quad x_1>0\,,
\\
{}& v^-_l\,,\quad x_1<0\,.
\end{aligned}\right.
\end{equation*}
The solution of (\ref{3.1}) is then constructed as an
interpolation (\ref{2.5}) with $v$ and $w$ replaced by $v_l$ and
$w_l$; this leads us to the desired operator equation,
\begin{equation}
g+T_4(\l)g+T_{11}(\l,l)=f\,.\label{3.5}
\end{equation}
Here $T_4$ is the operator appearing in (\ref{2.6}) and
$T_{11}(\l,l):L^2(\Pi_A)\to L^2(\Pi_A)$ is a compact linear
operator which is holomorphic in $\l$ and jointly continuous
w.r.t. $(\l,l)$ provided $\l\in\mathcal{D}_\d$, and
$l\in[l_0,+\infty]$. The norm of the last named operator is
exponentially small as $l\to+\infty$ uniformly in
$\l\in\mathcal{D}_\d$:
\begin{equation} \label{3.10}
\|T_{11}\|=\mathcal{O}(\Ex^{-2l\sqrt{1-\l}})\,.
\end{equation}
The solution to the problem (\ref{3.1}) can be reconstructed from
the function $g$ by $u=T_3(\l)g+T_{12}(\l,l)g$, where $T_{12}:
L^2(\Pi_A)\to\SS^1(\Pi^l)$ is a linear bounded operator the norm
of which satisfies
\begin{equation} \label{3.8}
\|T_{12}\|=\mathcal{O}(\Ex^{-l\sqrt{1-\l}})\,.
\end{equation}
This operator is also holomorphic in $\l$ and jointly continuous
with respect to $(\l,l)\in \mathcal{D}_\d\times[\l_0,+\infty]$.
The equation (\ref{3.5}) is a second-kind Fredholm operator
equation and it is equivalent to the problem (\ref{3.1}); this
claim can be checked in the same way as we did it for (\ref{2.6})
in the previous section.

The case of $hu=\frac{\p u}{\p x_1}$ is treated in full analogy.
The only difference due to another boundary condition at $x_1=0$
is the definition of the operator $T_{10}$ which is now described
by the kernel
\begin{equation}\label{3.6}
G^-_l(x,t,\l)=\sum\limits_{j=1}^\infty\frac{2\Ex^{-\kappa_j(\l)
l}}{\pi \kappa_j(\l)\cosh \kappa_j(\l) l}\, \sinh \kappa_j(\l)
x_1\, \sinh \kappa_j(\l) t_1\,\sin j x_2\,\sin j t_2\,.
\end{equation}
All the arguments used above remain valid.

On the other hand, for $\l$ in the vicinity of one almost all
the above arguments remain valid provided we replace $\l$ by
$(1-\kappa^2)$. In analogy with the previous section the
operators introduced here may be considered on cut-off strips,
i.e. as $T_9(\kappa,l):L^2(\Pi_A^-)\to \SS^2(\Pi^-_R)$,
$T_{10}(\kappa,l):L^2(\Pi_A^-)\to \SS^2(\Pi^-_R)$,
$T_{11}(\kappa,l):L^2(\Pi_A)\to L^2(\Pi_A)$, $T_{12}(\kappa,l):
L^2(\Pi_A)\to\SS^1(\Pi_R)$ for any fixed $R$. However, we are
not longer allowed to say that these operators are holomorphic
in $\kappa$ because of the terms
\begin{equation*}
\frac{2\Ex^{-\kappa l}}{\sinh \kappa l}\,,\quad
\frac{2\Ex^{-\kappa l}}{\cosh \kappa l}
\end{equation*}
in (\ref{3.7}), (\ref{3.6}), since these terms have poles at
$\kappa=\frac{\pi\mathrm{i}}{l} j$ and
$\kappa=\frac{\pi\mathrm{i}}{l} (j+\frac{1}{2})$. Moreover,
these terms are also responsible for the fact that the operators
have no proper limit as $\kappa\to0$ and $l\to+\infty$. At the
same time, restricting the range of $\kappa$ we will be able to
show that the operators $T_{10}$, $T_{11}$, and $T_{12}$ are
small for small $\kappa$ and large $l$, thus we will be allowed
to consider them as small perturbations again. This claim leans
on the following lemma.
\begin{lemma}\label{lm3.2}
Let $\k\in(0,\frac{\pi}{2})$ be fixed and $\mathcal{Q}_\k:=
\{\kappa: |\arg{\kappa}\pm\frac{\pi}{2}|\ge\k\}$. Then there is
$C>0$ such that for small $\kappa\in \mathcal{Q}_\k$ and large
$l$ the following estimate is valid,
\begin{equation*}
\max\left\{\left|\frac{\kappa\,\Ex^{-\kappa l}}{\sinh{\kappa
l}}\right|,\, \left|\frac{\kappa\,\Ex^{-\kappa l}}{\cosh{\kappa
l}}\right| \right\}\le C\left(|\kappa|+l^{-1}\right)\,.
\end{equation*}
\end{lemma}
\PF{Proof.} We will show how to derive the first estimate, the
proof of the second one is similar.  We start by introducing the
function
\begin{equation*}
P(z):=\frac{z}{\Ex^z-1}\,.
\end{equation*}
Suppose that $z\in\mathcal{Q}_\k$. If we have in addition
$|z|\le1$, one can check that
\begin{equation}\label{3.9}
|P(z)|\le C
\end{equation}
with some $C$ independent on $z$. On the other hand, if $|z|>1$,
$z\in\mathcal{Q}_\k$, and $\RE z>0$, then the exponent in the
function $P$ increases as $|z|\to\infty$ and we arrive at
(\ref{3.9}) again (in general with another $C$). Finally, if
$|z|>1$, $z\in\mathcal{Q}_\k$, and $\RE z<0$ then the exponent in
the function $P$ decreases and we have a uniform estimate,
\begin{equation*}
|P(z)|\le C|z|.
\end{equation*}
Combining it with (\ref{3.9}) we get the inequality
\begin{equation*}
|P(z)|\le C_1|z|+C_2
\end{equation*}
valid for $z\in\mathcal{Q}$ and suitable $C_1,C_2$. The obvious
identity
\begin{equation*}
\frac{\kappa\,\Ex^{-\kappa l}}{\sinh \kappa l}=\frac{1}{l}\,
P(2\kappa l)
\end{equation*}
then completes the proof of the lemma. \quad \QED

\medskip

Using this result one can check that the operators $T_{10}$,
$T_{11}$, and $T_{12}$ are small for small $\kappa\in
\mathcal{Q}_\k$ and large $l$, holomorphic in $\kappa$, and
jointly continuous in $(\kappa,l)$.

\section{Proof of Theorem~\ref{th1}}

In this section we are going to derive the asymptotic expansions
for the eigenvalues of $H_l(a)$ separated from the continuum. We
will also find the asymptotic behavior of the associated
eigenfunctions.

The main idea behind the calculation of the asymptotics is
borrowed from \cite{Ga1, Ga2, BEG}. Instead of dealing with
eigenvectors of $H_l(a)$ directly we consider here those of the
problems (\ref{1.2}). In order to find eigenvalues of the latter
we should look in accordance with the results of the previous
sections for $\l$ such that the operator equation
\begin{equation}\label{4.1}
\Phi+T_4(\l)\Phi+T_{11}(\l,l)\Phi=0
\end{equation}
has a nontrivial solution. We will deal with eigenvalues which are
close to a fixed eigenvalue $\l_j(a)$ of the limiting operator
$H(a)$; for simplicity we will denote the latter as $\l_0$ in the
following. Also the parameter $\l$ will be assumed to be close to
$\l_0$, more specifically, it will be supposed to lie in a
neighborhood of $\l_0$ containing neither any other limiting
eigenvalue nor the point $\l=1$.

By the definition of $T_{11}$ the term $T_{11}(\l,l)\Phi$ in
(\ref{4.1}) is supported inside $\Pi_A$. Hence considering it as
the right hand side, we arrive at the equation (\ref{2.6}) with
$f=-T_{11}(\l,l)\Phi$. Choosing $\l\not=\l_0$, we can invert the
operator $I+T_4(\l)$ obtaining
\begin{equation*}
\Phi+(I+T_4(\l))^{-1}T_{11}(\l,l)\Phi=0\,.
\end{equation*}
Using the Lemma~\ref{lm2.2}, we can rewrite the last equation in
the form
\begin{equation}\label{4.2}
\Phi-\frac{\phi}{\l-\l_0}(\psi,T_{11}(\l,l)\Phi)_{L^2(\Pi)}
+T_6(\l)T_{11}(\l,l)\Phi=0\,;
\end{equation}
recall that $\psi\in L^2(\Pi)$ here is the normalized
eigenfunction associated with $\l_0$ and $\phi\in L^2(\Pi_A)$ is a
function such that $\psi=T_3(\l_0)\phi$.

The operator $T_{11}(\l,l)$ is small in the asymptotic region,
$l\to+\infty$, while $T_6(\l)$ is holomorphic in $\l$. Thus we may
invert the operator $I+T_6(\l)T_{11}(\l,l)$ and apply the result
to the equation (\ref{4.2}), which then acquires the form
\begin{equation}\label{4.3}
\Phi-\frac{1}{\l-\l_0}(\psi,T_{11}(\l,l)\Phi)_{L^2(\Pi)}
(I+T_6(\l)T_{11}(\l,l))^{-1}\phi=0\,.
\end{equation}
The inner product $(\psi,T_{11}(\l,l)\Phi)_{L^2(\Pi)}$ does not
vanish. Indeed, otherwise the function $\Phi$ would be zero too,
however, we seek a nontrivial solution of the equation
(\ref{4.1}). With this fact in mind, we express the function
$\Phi$ from the equation (\ref{4.3}) and then calculate the inner
product $(\psi,T_{11}(\l,l)\Phi)_{L^2(\Pi)}$. This procedure leads
us to the equation
\begin{equation*}
1-\frac{1}{\l-\l_0}\left(\psi,T_{11}(\l,l)
(I+T_6(\l)T_{11}(\l,l))^{-1}\phi\right)_{L^2(\Pi)}=0\,,
\end{equation*}
or in a more convenient form
\begin{equation}\label{4.4}
\l-\l_0-\left(\psi,T_{11}(\l,l)
(I+T_6(\l)T_{11}(\l,l))^{-1}\phi\right)_{L^2(\Pi)}=0\,.
\end{equation}
This is the sought equation determining the perturbed eigenvalues
of the problem (\ref{1.2}), and, thus, of the operator $H_l(a)$.
The associated solution of the equation (\ref{4.1}), as it follows
from (\ref{4.3}), can be written as
\begin{equation}\label{4.5}
\Phi=(I+T_6(\l)T_{11}(\l,l))^{-1}\phi\,;
\end{equation}
we naturally keep in mind the fact that the eigenfunctions are
defined up to a multiplicative constant.

The equation (\ref{4.4}) determine all eigenvalues of $H_l(a)$;
due to the equivalence between (\ref{4.1}) and (\ref{1.2}) only
the eigenvalues of $H_l(a)$ satisfy this equation. Thus, by
Proposition~\ref{lmpr}, for every $T_{11}$ there exists an unique
solution of the equation (\ref{4.4}) converging to $\l_0$ as
$l\to+\infty$.

The desired asymptotic expansions for the perturbed eigenvalues
can be calculated directly from the equation (\ref{4.4}). First of
all we recall the assertion (\ref{3.10}) which implies that for
$\l$ close to $\l_0$ the norm $T_{11}$ can be estimated by
$\mathcal{O}(\Ex^{-(2\sqrt{1-\l_0}-\si)l})$. It allows us first to
establish the estimate
\begin{equation}\label{4.14}
\l-\l_0=\mathcal{O}\left(\Ex^{-(2\sqrt{1-\l_0}-\si)l}\right),
\end{equation}
and secondly to expand the second term in the equation (\ref{4.4})
obtaining
\begin{equation}\label{4.6}
\l-\l_0-\left(\psi,T_{11}(\l,l)\phi\right)_{L^2(\Pi)}
+\mathcal{O}(\Ex^{-2(2\sqrt{1-\l_0}-\si)l})=0\,.
\end{equation}
We can also extract the leading term from the operator
$T_{11}(\l,l)$, which obviously comes from the lowest-mode
contribution to the sum at the right hand side of (\ref{3.7}). We
will do that for $hu=u$, in the other case one proceeds
analogously.

First we introduce additional notations setting
\begin{align}\label{4.7}
V(x):=\left\{
\begin{aligned}
{}&-\frac{4\Ex^{-2\kappa_1(\l_0) l}}{\pi \kappa_1(\l_0)}\sinh
\kappa_1(\l_0) x_1\sin x_2\int\limits_{\Pi^-}\sinh \kappa_1(\l_0)
t_1\sin t_2\phi\,d^2t\,,{}& {}&x_1<0\,,
\\
{}&\hphantom{-\frac{4\Ex^{-2\kappa_1(\l) l}}{\pi
\kappa_1(\l)}\sinh \kappa_1(\l_0) x_1}0\,,{}& {}& x_1>0\,.
\end{aligned}\right.
\end{align}
Suppose that a function $W$ solves the problem (\ref{2.4}) with
$v=V$, then
\begin{equation*}
T_{11}(\l,l)\phi=-(\D+\l_0)\left(V+\chi(W\!-\!V)\right)
+\mathcal{O}(l\Ex^{-2(2\sqrt{1-\l_0}-\si)l})\quad \text{in
$L^2(\Pi_A)$}\,.
\end{equation*}
Using this identity together with the fact that the function
$T_{11}(\l,l)\phi$ is finite, we can calculate the leading term of
the second summand in (\ref{4.6}),
\begin{equation}
\begin{aligned}
{}&\left(\psi,T_{11}(\l,l)\phi\right)_{L^2(\Pi)}=-\int\limits_{\Pi}
(\D+\l_0)\left(V+\chi(W\!-\!V)\right)\,d^2x
+\mathcal{O}(l\Ex^{-2(2\sqrt{1-\l_0}-\si)l})\,,
\\
{}& \int\limits_{\Pi} (\D+\l_0)(V+\chi(W\!-\!V))\,d^2x=
\lim\limits_{R\to+\infty}\!\!
\int\limits_{\{x:\,|x_1|=R,\,0<x_2<\pi\}}\left( \psi\frac{\p V}{\p
\nu}- V\frac{\p \psi}{\p \nu}\right)\,ds
\\
{}&=-\lim\limits_{R\to+\infty}\int\limits_{\{x:\,x_1=-R,\,
0<x_2<\pi\}}\left( \psi\frac{\p V}{\p x_1}- V\frac{\p \psi}{\p
x_1}\right)\,ds\,.
\end{aligned}\label{4.8}
\end{equation}
In order to calculate the last integral we use the fact that in
view of the relation $\psi=T_3(\l_0)\phi$ and the definition of
$T_3$ the constant $\a=\a_j$ in (\ref{1.5}) is given by
\begin{align}
\a=-\frac{2\rho}{\pi \kappa_1(\l_0)}
\int\limits_{\Pi^-}\sinh(\kappa_1(\l_0) t_1)\sin t_2\,
\phi(t)\,d^2t\,,\nonumber
\end{align}
where $\rho$ is 1 if $\psi$ even and -1 if it is odd. Using this
relation together with (\ref{1.5}) and (\ref{4.7}), we can finish
our calculations in (\ref{4.8}) arriving at
\begin{equation*}
\lim\limits_{R\to+\infty}\!\!
\int\limits_{\{x:\,x_1=-R,\,0<x_2<\pi\}}\left( \psi\frac{\p V}{\p
x_1}- V\frac{\p \psi}{\p x_1}\right)\,ds =\pi
\a^2\kappa_1(\l_0)\,\Ex^{-2\kappa_1(\l_0)l}\,.
\end{equation*}
Combining this with (\ref{4.8}) and (\ref{4.6}) we get the
asymptotic (\ref{asm}), (\ref{ltd1}) for $\l_j^-(a)$. In the case
$hu=\frac{\p u}{\p x_1}$ a similar reasoning leads to asymptotics
(\ref{asm}), (\ref{ltd1}) for $\l_j^+(a)$. In order to prove
relation (\ref{ldt}) it is sufficient to express $\a$ in terms of
suitable integrals. Keeping the parity of $\psi$ in mind we
compute
\begin{align*}
0{}&=\lim\limits_{R\to+\infty}\int\limits_{\Pi_R}\Ex^{\kappa_1(\l_0)
x_1}\sin x_2\,(\D\!+\!\l_0)\psi(x)\,d^2x
\\
{}&=\lim\limits_{R\to+\infty}\int\limits_{\p\Pi_R}\left(\Ex^{
\kappa_1(\l_0) x_1}\sin x_2\frac{\p}{\p\nu}\psi(x)
-\psi(x)\frac{\p}{\p\nu}\,\Ex^{\kappa_1(\l_0) x_1}\sin
x_2\right)\,ds
\\
{}&=\int\limits_{\g(a)}\psi(x)\,
\Ex^{\sqrt{1-\l_0}x_1}\,dx_1-\a\pi\sqrt{1\!-\!\l_0}\,.
\end{align*}
This result leads us to formulae (\ref{ldt}).

The asymptotics of the eigenfunctions can be derived easily. The
definite parity of those associated with $\l_j^\pm(l,a)$ is
obvious. The relation (\ref{4.5}) tells us that
\begin{equation}\label{4.12}
\Phi^\pm=\phi+\mathcal{O}(\Ex^{-(l\sqrt{1-\l_0}-\si)})\,.
\end{equation}
The symbol ''$\pm$'' indicate here two variants of definition of
the operator $T_{11}$. Now in order to prove the expansions for
the eigenfunctions one has just to use this expression and to
employ the arguments of the previous two section. More precisely,
we have $\psi=T_3(\l_0)\phi$ and
$\Psi^\pm=(T_3(\l^\pm)+T_{12}(\l^\pm,l))\Phi^\pm$, where
$\Psi^\pm$ is the eigenfunction of the problem (\ref{1.2})
associated with the chosen eigenvalue and chosen variant of
boundary operator $h$. Using (\ref{4.12}) and the holomorphy of
$T_3$, $T_{12}$, the estimates (\ref{4.14}) and (\ref{3.8}), we
arrive at the asymptotical formula
\begin{equation}\label{4.15}
\Psi^\pm=\phi+\mathcal{O}\left(\Ex^{-l(\sqrt{1-\l_0}-\si)l}\right)
\end{equation}
in $\SS^1(\Pi^l)$. Recovering now the eigenfunctions of $H_l(a)$
we obtain all their properties stated in Theorem~\ref{th1}.

Let us finally prove that there are no other eigenvalues of
$H_l(a)$ in $\mathcal{D}_1$. Consider the equation (\ref{4.1})
where $\l$ is close to one and does not lie in  real semi-axis
$[1,+\infty)$, more specifically, suppose that $\kappa\in
\mathcal{Q}_\k$. Then we can invert the operator
$(I+T_4(\kappa))^{-1}$, and arrive at the equation
\begin{equation*}
\Phi+(I+T_4(\kappa))^{-1}T_{11}(\kappa,l)\Phi=0\,,
\end{equation*}
where the operator $(I+T_4(\kappa))^{-1}$ is uniformly bounded in
$\kappa$, because the Neumann segment does not have by assumption
a critical size -- see Lemma~\ref{lm2.3} -- while
$T_{11}(\kappa,l)$ is small for all possible values $\kappa$ and
$l$. Hence the operator $(I+T_4(\kappa))^{-1}T_{11}(\kappa,l)$ is
also small, and therefore we can invert in turn the operator
$(I+(I+T_4(\kappa))^{-1}T_{11}(\kappa,l))$ which immediately leads
us to the unique solution $\Phi=0$. Moreover, the operator
$H_l(a)$ cannot have eigenvalues corresponding to $\kappa$
satisfying $|\arg \kappa\pm\frac{\pi}{2}|<\k$, $\RE\kappa\not=0$,
simply because it is self-adjoint and all its eigenvalues are
real, thus there is no other eigenvalues to $H_l(a)$ in
$\mathcal{D}_1$. By this the proof of Theorem~\ref{th1} is
complete. \quad \QED

\section{Proof of Theorem~\ref{th3}}

It is sufficient to consider in detail only the eigenvalue
$\l_{n+1}$ emerging from the continuum because all the
statements related to the other eigenvalues verify in a way
completely analogous to the previous section.

We know from Proposition~\ref{lmpr} that the eigenfunction
associated with the indicated eigenvalue is even with respect to
$x_1$, thus we have to consider here only the case $hu=\frac{\p
u}{\p x_1}$. Assuming $\kappa\in\mathcal{Q}_\k$, we start with the
equation
\begin{equation}\label{6.0}
\Phi+T_4(\kappa)\Phi+T_{13}(\kappa,l)\Phi=0\,,
\end{equation}
which is how (\ref{4.1}) looks like in the present case, with
$T_{13}(\kappa,l)$ being the perturbation operator associated with
(\ref{3.6}). The operator $T_{13}(\kappa,l)$ is small by
Lemma~\ref{lm3.2}, and an argument analogous to that which lead us
to (\ref{4.3}) yields the equation
\begin{equation}\label{6.1}
\Phi+\frac{1}{2\kappa}(\psi,T_{13}(\kappa,l)\Phi)_{L^2(\Pi)}
(I+T_8(\l)T_{13}(\l,l))^{-1}\phi=0\,.
\end{equation}
Recall that $\psi=\psi^n$ is a solution to the equation
$(H(a_n)\!+\!1)\psi^n=0$ which behaves at infinity in accordance
with (\ref{1.1}) and $\phi\in L^2(\Pi_A)$ such that
$\psi=T_3(\kappa=0)\phi$. From this equation one can deduce an
analogue of the equation (\ref{4.4}), namely
\begin{equation}\label{6.2}
2\kappa+\left(\psi,T_{13}(\kappa,l)(I+T_8(\kappa)
T_{13}(\kappa,l))^{-1}\phi\right)_{L^2(\Pi)}=0\,.
\end{equation}
The value of $\kappa$ associated with the eigenvalue emerging from
the continuum solves this equation and by Proposition~\ref{lmpr}
it tends to zero. Using these two facts we will deduce the
asymptotic formula stated in Theorem~\ref{th3}. First of all, in
the following we will consider the equation (\ref{6.2}) for real
positive $\kappa$ only. This restriction can be justified easily,
since for negative $\kappa$ the associated function $u$ given by
(\ref{2.5}) increases at infinity and thus it does not belong to
$L^2(\Pi)$. In order to calculate the asymptotics, we extract the
leading part of the second term in the equation (\ref{6.2}); for
small positive $\kappa$ we have
\begin{equation}\label{6.3}
\left(\psi,T_{13}(\kappa,l)(I\!+\!T_8(\kappa)T_{13}
(\kappa,l))^{-1}\phi\right)_{L^2(\Pi)}=
\left(\psi,T_{13}(\kappa,l)\phi\right)_{L^2(\Pi)}
+T_{14}(\kappa,l)\,,
\end{equation}
where $T_{13}(\kappa,l):\mathbb{R}^2\to\mathbb{R}$ is a function
defined for $(\kappa,l)\in\mathcal{Q}_\k\times[\l_0,+\infty)$
which satisfies the relation
\begin{equation}\label{6.5}
T_{14}(\kappa,l)=\mathcal{O} \left(\frac{\kappa^2\Ex^{-2\kappa
l}}{\cosh^2 \kappa l}+\Ex^{-4\sqrt{3}l}\right)
\end{equation}
as $(\kappa,l)\to(0,+\infty)$. To get this estimate one has to
employ the relation
$T_{14}(\kappa,l)=\mathcal{O}(\|T_{13}(\kappa,l)\|^2)$ and the
fact that
\begin{equation}\label{6.12}
\|T_{13}(\kappa,l)\|=\mathcal{O}\left(\frac{\kappa\,
\Ex^{-\kappa l}}{\cosh \kappa l}+\Ex^{-2\sqrt{3}l}\right)
\end{equation}
implied by the definition of $T_{13}$ -- see (\ref{3.6}). Our next
step is to extract the leading term from the first summand at the
right hand side of (\ref{6.3}). We will do it in the same way as
in last section, the only difference is that now we have to take
into account also the second transverse-mode contribution to
(\ref{3.6}).

We introduce the function $V_1$ that is an analogue of (\ref{4.7})
by
\begin{align}\label{6.4}
V_1(x):=\left\{
\begin{aligned}
{}&\frac{2\kappa\,\Ex^{-\kappa l}}{\pi \cosh \kappa l}\, x_1
\sin x_2\int\limits_{\Pi^-} t_1\sin t_2\,\phi(t)\,d^2t\,,{}&
{}&x_1<0\,,
\\
{}&\hphantom{\frac{2m\Ex^{-m l}}{\pi \cosh m l} x_1 \sin
x_2}0\,,{}& {}& x_1>0\,.
\end{aligned}\right.
\end{align}
Let $W_1$ be a solution to the problem (\ref{2.4}) with $v=V_1$.
We also introduce the function $V_2$ in the following way
\begin{align}\label{6.11}
V_2(x):=\left\{
\begin{aligned}
{}&\frac{2\Ex^{-\sqrt{3}l}}{\pi\sqrt{3}\cosh \sqrt{3} l}
\sinh\sqrt{3}x_1 \sin 2x_2\int\limits_{\Pi^-}
\sinh\sqrt{3}t_1\sin 2t_2\,\phi(t)\,d^2t\,,{}& {}&x_1<0\,,
\\
{}&\hphantom{\frac{2m\Ex^{-m l}}{\pi \sinh m l} x_1 \sin
x_2}0\,,{}& {}& x_1>0
\end{aligned}\right.
\end{align}
and suppose that $W_2$ is a solution of (\ref{2.4}) with $v=V_2$.
One can check that
\begin{equation*}
(\psi,T_{13}(\kappa,l)\phi)_{L^2(\Pi)}=-\left(\psi,
(\D\!+\!1)\left(\t V+\chi(\t W\!-\!\t
V)\right)\right)_{L^2(\Pi)}+T_{15}(\kappa,l)\,,
\end{equation*}
where $\t V=V_1+V_2$, $\t W=W_1+W_2$, and the function
$T_{15}(\kappa,l)$ satisfies the estimate
\begin{equation}\label{6.7}
T_{15}(\kappa,l)=\mathcal{O}\left(\frac{\kappa^3\Ex^{-\kappa
l}}{\cosh \kappa
l}+\kappa^2\Ex^{-2\sqrt{3}l}+\Ex^{-2\sqrt{8}l}\right)
\end{equation}
as $(\kappa,l)\to(0,+\infty)$. Calculating the inner product
$\left(\psi,T_{13}(\kappa,l)\phi\right)_{L^2(\Pi)}$ in the same
way how we deduced (\ref{asm}) and bearing in mind the asymptotics
(\ref{1.1}) for $\psi$ together with (\ref{6.3}), (\ref{6.5}), and
(\ref{6.4})--(\ref{6.7}) we obtain the equation
\begin{equation}\label{6.8}
\begin{aligned}
2\kappa+&\rho\frac{\sqrt{2}\kappa\Ex^{-\kappa l}}{\sqrt{\pi}\cosh
\kappa l} \int\limits_{\Pi^-} t_1\sin t_2\,\phi\,d^2t
\\
+&\rho
\frac{\b\,\Ex^{-\sqrt{3}l}}{\cosh\sqrt{3}l}\int\limits_{\Pi^-}
\sinh\sqrt{3}t_1\sin 2t_2\,\phi\,d^2t+T_{16}(\kappa,l)=0\,,
\end{aligned}
\end{equation}
where $\rho$ is again the parity of $\psi$. The function
$T_{16}(\kappa,l)$ satisfies
\begin{equation}\label{6.6}
T_{16}(\kappa,l)=O\left(\frac{\kappa^2\Ex^{-2\kappa l}}{\cosh^2
\kappa l}+\frac{\kappa^3\Ex^{-\kappa l}}{\cosh \kappa
l}+\kappa^2\Ex^{-2\sqrt{3}l}+\Ex^{-2\sqrt{8}l}\right)
\end{equation}
as $(\kappa,l)\to(0,+\infty)$. Since the function $\phi$ obeys
$\psi=T_3(\kappa=0)\phi$, we can take into account the definition
of the last operator (see (\ref{2.13})) and the asymptotics
(\ref{1.1}) to conclude that
\begin{align*}
{}&\rho\sqrt{\frac{2}{\pi}}=-\frac{2}{\pi}\int\limits_{\Pi^-}
t_1\sin t_2\,\phi(t)\,d^2t\,,
\\
{}&\rho\b=-\frac{2}{\pi\sqrt{3}}\int\limits_{\Pi^-} \sinh
\sqrt{3}t_1\sin 2t_2\,\phi(t)\,d^2t\,,
\end{align*}
which together with (\ref{6.8}) leads us to ($\b=\b_n$)
\begin{equation*}
2\kappa-\frac{\kappa\,\Ex^{-\kappa l}}{\cosh \kappa
l}-\frac{\b^2\pi\sqrt{3}}{2}\frac{\Ex^{-\sqrt{3}l}}{\cosh\sqrt{3}l}
+T_{16}(\kappa,l)=0\,,
\end{equation*}
or equivalently,
\begin{equation}\label{6.10}
\frac{\kappa\,\Ex^{\kappa l}}{\cosh \kappa
l}-\frac{\b^2\pi\sqrt{3}}{2}\frac{\Ex^{-\sqrt{3}l}}{\cosh\sqrt{3}l}
+T_{16}(\kappa,l)=0\,.
\end{equation}
We know that this equation has a positive solution tending to zero
as $\kappa\to0$. In view of (\ref{6.6}), (\ref{6.10}), and the
trivial inequality
\begin{equation*}
1\le\frac{\Ex^{\tau}}{\cosh\tau}\le2
\end{equation*}
we have for this solution the following estimate,
\begin{equation*}
C_1\Ex^{-2\sqrt{3}l} \le \kappa\le C_2\Ex^{-2\sqrt{3}l}
\end{equation*}
with constants $C_1$, $C_2$ independent on $l$.
Using it we can expand the first term in the equation (\ref{6.10})
with respect to $\kappa l$, which is small, and to estimate
$T_{15}$ -- see (\ref{6.6}). In this way we arrive at the
relation,
\begin{equation*}
\kappa-\b^2\pi\sqrt{3}\,\Ex^{-2\sqrt{3}l}
+\mathcal{O}(\Ex^{-2\sqrt{8}l})=0\,,
\end{equation*}
which implies the sought asymptotical expansion (\ref{1.3}),
(\ref{1.4}). The second formula for $\mu$ stated in
Theorem~\ref{th3} can be proven completely by analogy with the
proof of (\ref{ldt}). One just should multiply the equation
$(\Delta+1)\psi^n$ by $\Ex^{\sqrt{3}x_1}\sin 2x_2$ and integrate
then by parts over $\Pi_R$ passing then to the limit as
$R\to+\infty$.

The argument concerning the asymptotics for the associated
eigenfunction is completely analogous to that of the previous
section. The solution to the equation (\ref{6.0}) is given by
\begin{equation*}
\Phi=(I+T_8(\kappa)T_{13}(\kappa,l))^{-1}\phi\,.
\end{equation*}
Now one has just to perform the expansion using the fact that the
operator $T_8(\kappa)T_{13}(\kappa,l)$ is small, then using the
obtained asymptotics for $\kappa$, to apply to the remainder the
estimate (\ref{6.12}),  to construct the corresponding
eigenfunction of the problem (\ref{1.2}) by the scheme described
in the Section~5, and finally,  to recover the eigenfunctions of
$H_l(a_n)$. This completes the proof of the second theorem. \quad
\QED


\subsection*{Acknowledgments}

D.B. is grateful for the hospitality in the Department of
Theoretical Physics, NPI, Czech Academy of Sciences, where a part
of this work was done. The research has been partially supported
by GAAS under the contract A1048101, by RFBR under the contracts
02-01-00693, 03-01-06470 and by the program ''Leading scientific
schools'' (NSh-1446.2003.1).

\end{document}